\documentclass[fleqn,twoside]{article}
\usepackage{espcrc2,psfig}

\newlength{\figwidth}
\renewcommand{\vec}[1]{\mbox{\boldmath $#1$}}

\title{Heavy-light meson in anisotropic lattice QCD%
       \thanks{Talk presented by H. Matsufuru}}

\author{
H. Matsufuru\address{%
    Yukawa Institute for Theoretical Physics, Kyoto University,
    Kyoto 606-8502, Japan \vspace{-0.25cm} },
J. Harada\address{%
    Department of Physics, Hiroshima University,
    Higashi-hiroshima 739-8526, Japan \vspace{-0.25cm}},
T.~Onogi$^{\rm a}$
and
A. Sugita$^{\rm a}$ }

\begin{document}

\begin{abstract}
We examine whether the $O(a)$ improved quark action on anisotropic
lattices can be used as a framework for the heavy quark,
which enables precision computation of matrix
elements of heavy-light mesons.
To this end, it is crucial to verify that a mass independent and
nonperturbative tuning of the parameters is possible.
As a first step, we observe the dispersion relation of heavy-light
mesons on a quenched lattice using the action
which is nonperturbatively tuned only for the leading terms.
On a lattice with the spatial cutoff $a_\sigma^{-1} \simeq$ 1.6 GeV
and the anisotropy $\xi=4$,
the relativity relation holds within 2\% accuracy in the quark
mass region $a_\sigma m_Q \leq 1.2$
with the bare anisotropy parameter tuned for the massless quark.
We also apply the action to a calculation of
heavy-light decay constants in the charm quark mass region.
\end{abstract}

\maketitle

\section{Introduction}
  \label{sec:introduction}

Recent experimental progress at $B$ factories 
suggests that precise computation of hadronic 
matrix elements in lattice QCD is a key 
to the search for signals of new physics in flavor physics.

However, the two major formalisms for the heavy quark, {\it i.e.}
HQET approaches or the relativistic approaches to 
describe the heavy quarks on the lattice still suffer 
from perturbative and/or discretization errors, 
which are typically $\sim$ 10\%.
We therefore need yet another formalism of the heavy 
quark for the precise and systematic computation of matrix elements.

Ideally, in such a formalism, one should be able to
(i) take the continuum limit,
(ii) compute the parameters in the action and the 
     operators nonperturbatively,
(iii) and compute  matrix elements with a modest  computational
      cost.

As a candidate of framework which fulfills these conditions,
we investigate the anisotropic lattice with the temporal lattice
spacing $a_\tau$ finer than the spatial one $a_\sigma$
\cite{Aniso01a,Aniso01b,Aniso02a}.
Anisotropic lattice framework evidently satisfies above conditions
(i) and (iii).
Our expectation is that on anisotropic lattices the mass dependence
of the parameters becomes so mild that one can adopt
coefficients determined nonperturbatively at massless limit.
Whether these promises will be practically satisfied should be
examined numerically.

As a first step toward this goal, we study the action
which is nonperturbatively tuned only for the leading term,
while $O(a)$ improvement terms are tuned at 
the tadpole tree level.
We measure the heavy-light meson dispersion
relation with the anisotropy parameter tuned at the massless limit.
We can then study the breaking of relativity condition 
as a function of the heavy quark mass, with which we can test 
if the small mass dependence of the parameter really holds.

\section{Anisotropic lattice quark action}
\label{sec:formulation}
Our heavy quark formalism  basically follows the
Fermilab approach \cite{EKM97} but is formulated on the anisotropic
lattice~\cite{Aniso01a,Ume01}.

The quark action is represented as
\begin{eqnarray}
 S_F &=& \sum_{x,y} \bar{\psi}(x) K(x,y) \psi(y),\\
 K(x,y) \!\!\!&=&\!\!\!
 \delta_{x,y}
   - \kappa_{\tau} \left[ \ \ (1-\gamma_4)U_4(x)\delta_{x+\hat{4},y} \right.
 \nonumber \\
 & &  \hspace{1cm}
      + \left. (1+\gamma_4)U_4^{\dag}(x-\hat{4})\delta_{x-\hat{4},y} \right]
 \nonumber \\
 & & \hspace{-0.2cm}
    -  \kappa_{\sigma} {\textstyle \sum_{i}}
         \left[ \ \ (r-\gamma_i) U_i(x) \delta_{x+\hat{i},y} \right.
 \nonumber \\
 & & \hspace{1cm}
     + \left. (r+\gamma_i)U_i^{\dag}(x-\hat{i})\delta_{x-\hat{i},y} \right]
 \nonumber \\
 & & \hspace{-0.2cm}
    -  \kappa_{\sigma} c_E
             {\textstyle \sum_{i}} \sigma_{4i}F_{4i}(x)\delta_{x,y}
 \nonumber \\
 & & \hspace{-0.2cm}
    - r \kappa_{\sigma} c_B
             {\textstyle \sum_{i>j}} \sigma_{ij}F_{ij}(x)\delta_{x,y},
 \label{eq:action}
\end{eqnarray}
where $\kappa_{\sigma}$ and  $\kappa_{\tau}$ 
are the spatial and temporal hopping parameters, $r$ is the Wilson
parameter and  $c_E$ and $c_B$ are  the clover coefficients.

In principle for a given $\kappa_{\sigma}$, the 
four parameters $\kappa_{\sigma}/\kappa_{\tau}$, $r$, $c_E$ and $c_B$
should be tuned so that Lorentz symmetry holds up to 
discretization errors of $O(a^2)$.

In this work, we set the spatial Wilson parameter as $r=1/\xi$ and
the clover coefficients as the tadpole-improved tree-level
values as $c_E= 1/u_{\sigma} u_{\tau}^2$, $  c_B = 1/u_{\sigma}^3$
and perform a nonperturbative calibration only for $\gamma_F$.
The tadpole improvement \cite{LM93} is achieved
by rescaling the link variables as
$U_i(x) \rightarrow U_i(x)/u_{\sigma}$ and  $U_4(x) \rightarrow
U_4(x)/u_{\tau}$, with the mean-field values of the spatial 
and temporal link variables, $u_{\sigma}$ and $u_{\tau}$,
respectively.
This is equivalent to redefining the
hopping parameters with the tadpole-improved ones (with tilde)
through $\kappa_{\sigma} = \tilde{\kappa}_{\sigma}/u_{\sigma}$
and $\kappa_{\tau} = \tilde{\kappa}_{\tau}/u_{\tau}$.
We define the anisotropy parameter $\gamma_F$ as
$\gamma_F \equiv \tilde{\kappa}_{\tau}/\tilde{\kappa}_{\sigma}$.

For convenience, we also introduce $\kappa$ as
\begin{eqnarray}
\frac{1}{\kappa} \equiv \frac{1}{\tilde{\kappa}_{\sigma}}
     - 2(\gamma_F+3r-4)
\hspace{0.3cm}  = 2(m_{0}\gamma_F +4) ,
 \label{eq:kappa}
\end{eqnarray}
where $m_0$ is the bare quark mass in temporal lattice units.
$\gamma_F$ is the bare anisotropy parameter, which should be tuned
so that the physical fermionic anisotropy $\xi_F$ retains the
same value as that of the gauge field, $\xi_G$.
For the quenched case, as for the present work, one can first
calibrate the anisotropy parameter of gauge field separately from
the quark field, and then tune $\gamma_F$ so that $\xi_F=\xi_G=\xi$
holds.

The calibration of bare anisotropy $\gamma_F$ was performed in
Ref.~\cite{Aniso01b} using the dispersion relation of mesons
for quark masses less than the charm quark mass region,
on quenched lattices with $\xi=4$ for three different
lattice spacings.
The quark mass dependence is indeed mild and well represented
by a linear form in $m_q^2$.

\section{Relativity condition}
\label{sec:numerical}

We next study whether the relativistic dispersion relation holds
for the heavy-light mesons with $\gamma_F$ tuned for the
massless quark \cite{Aniso02a}.
In general, the lattice dispersion relation for arbitrary values 
of $\gamma_F$ is described as
\begin{equation}
E^2 (\vec{p}) = m^2 + \vec{p}^2/ \xi_F^2  + O(p^4),
\label{eq:DR1}
\end{equation}
where the energy $E$ and the rest mass $m$ 
are in temporal lattice units while the momentum $\vec{p}$ is
in spatial units.
The parameter $\xi_F$ in this equation defines
the anisotropy of the quark fields, whose difference with 
the gauge field anisotropy $\xi$ probes the breaking of relativity.

Numerical simulations are performed on a quenched lattice
of the size $16^3\times 128$ generated with the Wilson plaquette
action at $\beta=6.10$ for $\xi=4$, which is realized 
by taking the bare anisotropy of gauge field $\gamma_G$ 
using the numerical formula determined 
by Klassen in one percent accuracy \cite{Kla98}.
The lattice cutoff defined through the hadronic radius $r_0$
is $a_\sigma^{-1}=1.623(9)$ GeV.
Other details of parameters such as the mean-field values are found in
Refs.~\cite{Aniso01b,Aniso02a}.

\begin{figure}[tb]
\vspace*{-0.2cm}
\psfig{file=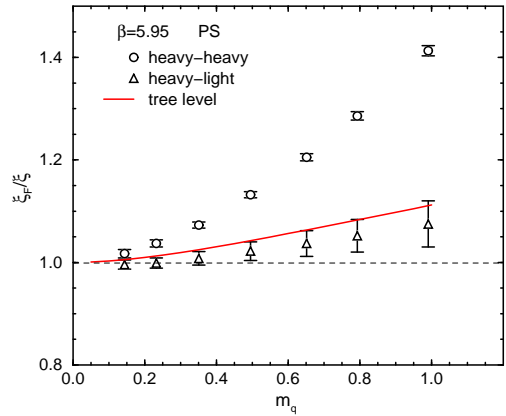,width=\figwidth}
\vspace{-1.2cm}
\caption{The heavy quark mass dependence of the anisotropy 
$\xi_F$ for heavy-light and heavy-heavy mesons
at $\beta=5.95$.}
\label{fig:xi_F}
\vspace{-0.4cm}
\end{figure}

The result for the pseudoscalar meson is displayed in Fig.~\ref{fig:xi_F}.
Similar result is obtained for the vector channel.
The measured value $\xi_F$ for the heavy-light meson coincides with $\xi$
in the quark mass region $m_q\leq 0.3$ (in temporal lattice units)
within 2\% accuracy.
(This accuracy is just a guide for argument of uncertainty in
this paper.)
Therefore, for the range of quark mass $m_q\leq 0.3$
which covers the charm quark mass
with present lattice cutoff, our quark action with the bare anisotropy 
tuned for massless quark can successfully describe the heavy-light 
meson with controlled systematic uncertainty.
We also show the $\xi_F$ of the heavy quarkonium.
In this case, $\xi_F$ rapidly increases
because of large momentum transfer inside heavy quarkonium,
$p\simeq \alpha m_Q$, in contrast to the case of heavy-light systems,
where $p\simeq \Lambda_{QCD}$.

\section{Heavy-light decay constant}
\label{sec:decayconst}

\begin{figure}[tb]
\vspace*{-0.2cm}
\psfig{file=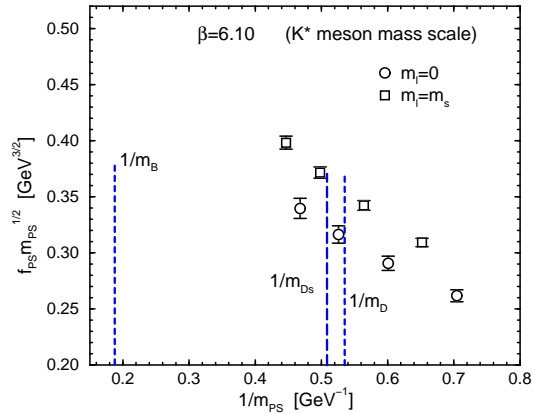,width=\figwidth}
\vspace{-1.2cm}
\caption{Heavy-light meson decay constant in the pseudoscalar
channel at $\beta=6.10$.}
\label{fig:decayconst}
\vspace{-0.4cm}
\end{figure}

The result in the previous section tells us that the framework can be
applied to charmed heavy-light meson systems.
As a first application, we compute the heavy-light decay constants.
Numerical simulations are performed on two quenched lattices
at $\beta=5.95$ and $6.10$ with the renormalized anisotropy $\xi=4$.
The former is the same as in previous section, and the latter
has spatial cutoff $a_\sigma^{-1}=2.030(13)$ GeV.
For the light quark, we use three values of the hopping parameter 
$\kappa$ which cover the range of quark masses
from $m_s$ to 1.5 $m_s$. The chiral limit is taken by a linear extrapolation.
The matching to the continuum theory is at the tadpole improved
tree level.
The result at $\beta=6.0$ is displayed in Fig.~\ref{fig:decayconst}. 
In this analysis, the physical quark masses are defined through
the lattice cutoff defined with the $K^*$ meson mass.
We find that the behavior of $f\sqrt{m}$ is consistent with the 
results on isotropic lattices.
Since the currents are matched only at the tadpole tree level
and the $O(a)$ improvement terms are tuned also only at the tadpole
tree level, the present calculation contains rather large renormalization
uncertainty as well as the cutoff dependence.
In order to suppress these errors we compute the ratio of decay
constants. Our preliminary results are 
\begin{eqnarray}
 f_D/f_{\pi} &=& 1.566(43) \mbox{\ \ at $\beta=5.95$}, \nonumber\\
             & & 1.515(43) \mbox{\ \ at $\beta=6.10$}, \nonumber\\
 f_{D_s}/f_{D} &=& 1.140(14) \mbox{\ \ at $\beta=5.95$}, \nonumber\\
               & & 1.142(14) \mbox{\ \ at $\beta=6.10$},
\end{eqnarray}
which are consistent with previous works on isotropic lattices
\cite{Ryan02}.
A remarkable point is that the lattice spacing dependence
of the ratio is small. In fact, individual decay constants 
could suffer from both the mass independent and mass dependent 
errors, but the former
can largely cancel in the ratio.
Therefore, the stability
of the ratio may be an indication that the mass dependent 
errors are suppressed in our formalism on anisotropic lattices
as expected.

In conclusion, the results of numerical
simulations in quenched
lattices are encouraging for further development in this direction.
Further improvements, such as nonperturbative tuning of the clover
coefficients, are necessary for achieving the desired accuracy.

The simulations were done on a NEC SX-5 at RCNP, Osaka University
and a Hitachi SR8000 at KEK.

\end{document}